\documentclass[%
 reprint,
 amsmath,amssymb,
 aps,
]{revtex4-2}

\usepackage{graphicx}% Include figure files
\usepackage{dcolumn}% Align table columns on decimal point
\usepackage{bm}% bold math
\usepackage{hyperref}% add hypertext capabilities
\begin{document}

%\preprint{APS/123-QED}

\title{Laser scattering by submicron droplets originated during the electrical explosion of thin metal wires}% Force line breaks with \\

\author{V.~M.~Romanova$^1$}
 %\email{romanovavm@lebedev.ru}
\author{G.~V.~Ivanenkov$^1$}%
\author{E.~V.~Parkevich$^1$}%
 \email{parkevich@phystech.edu}
\author{I.~N.~Tilikin$^1$}%
\author{M.~A.~Medvedev$^1$}%
\author{T.~A.~Shelkovenko$^1$}%
\author{S.~A.~Pikuz$^1$}%
\author{A.~S.~Selyukov$^{1,2,3,4}$}%

\affiliation{ $^1$ P. N. Lebedev Physical Institute of the Russian Academy of Sciences,\\ 53 Leninskiy pr., 119991 Moscow, Russia}
\affiliation{ $^2$ N. E. Bauman Moscow State Technical University,\\ 5/1 2-ya Baumanskaya St., 105005 Moscow, Russia}
\affiliation{ $^3$ All-Russian Institute for Scientic and Technical Information (VINITI)\\ of the Russian Academy of Sciences,\\ 20 Usievicha St., 125190 Moscow, Russia}
\affiliation{ $^4$ Moscow Institute of Physics and Technology (National Research University),\\ 9 Institutskiy Per., Dolgopruny 141700, Moscow Region, Russia}


\begin{abstract}
This paper presents the results of studying of dispersed media formation during the electrical explosion of thin metal wires in vacuum by using low-current generators ($\sim 1 \textrm{--} 10$~kA). Particular attention is paid to the analysis of the composition and structure of the corresponding explosion products as well as to the problem of their visualization using simultaneous laser interferometry and shadow imaging at two wavelengths (1.064~$\mu$m and 0.532~$\mu$m). Our findings point to the fact that the important role in the visualization of the explosion products belongs to multiple scattering by submicron droplets of dense condensed matter, which are mixed with metal vapor. The hypothesis on the existence of submicron droplets in the products of exploding metal wires correlates with the results obtained by soft x-ray radiography combined with a laser probing technique. Taking into account the multiple scattering by submicron droplets, it is possible to significantly clarify the parameters of the explosion products visualized via laser probing techniques as well as to gain a deeper insight into the physics behind the electrical wire explosion.
\end{abstract}

\maketitle

%\tableofcontents

\section{\label{sec1}Introduction}

Investigations of an electrical wire explosion (EWE) have great importance for fundamental physics of condensed matter in extreme states~\cite{clerouin_2008, stephens_2012} and multiple applications, e.g. soft x-ray radiography~\cite{shelkovenko_2016}, nanopowder manufacturing~\cite{bora_2014, pervikov_2019}, underwater shockwave production~\cite{gurovich_2004, grinenko_2005, krasik_2006, han_2017}, current switching~\cite{chuvatin_2006}, and many others. Basically, the wire explosion in vacuum, as well as in liquid or air media is understood as a complex transformation of the metal at a high current density into different forms of matter which, in particular, are metal vapor and foam, plasma, fine-scale droplets, and liquid microdrops. However, in spite of the fact that the EWE has been studied in various laboratories throughout the world for a long time, many aspects of the EWE remain veiled in mystery. Also, there is still no comprehensive theory describing the basic states of matter at different stages of the EWE due to the exceptional complexity and diversity of the processes involved in the generation of the explosion products. Therefore, the conclusion about one or another phase state attributed to the explosion products is mostly hypothetical. It should be noted that the explosion products are far from always amenable to direct diagnosing, and if they are, then the interpretation of the corresponding results can be extremely ambiguous.

The present work clearly demonstrates this problem by the example of the analysis of the results obtained when probing an exploding metal wire with two-wavelength laser pulses. In contrast to the established concepts explaining the possible principles of visualization of the explosion products generated during the EWE, we point out that here the key role belongs to the multiple scattering of the probed laser radiation by submicron droplets. The latter are assumed to be dense condensed matter generated at the early stage of the wire explosion.

\section{\label{sec2}Challenges in diagnosing explosion products}

\begin{figure*}
\includegraphics [width=15.6cm] {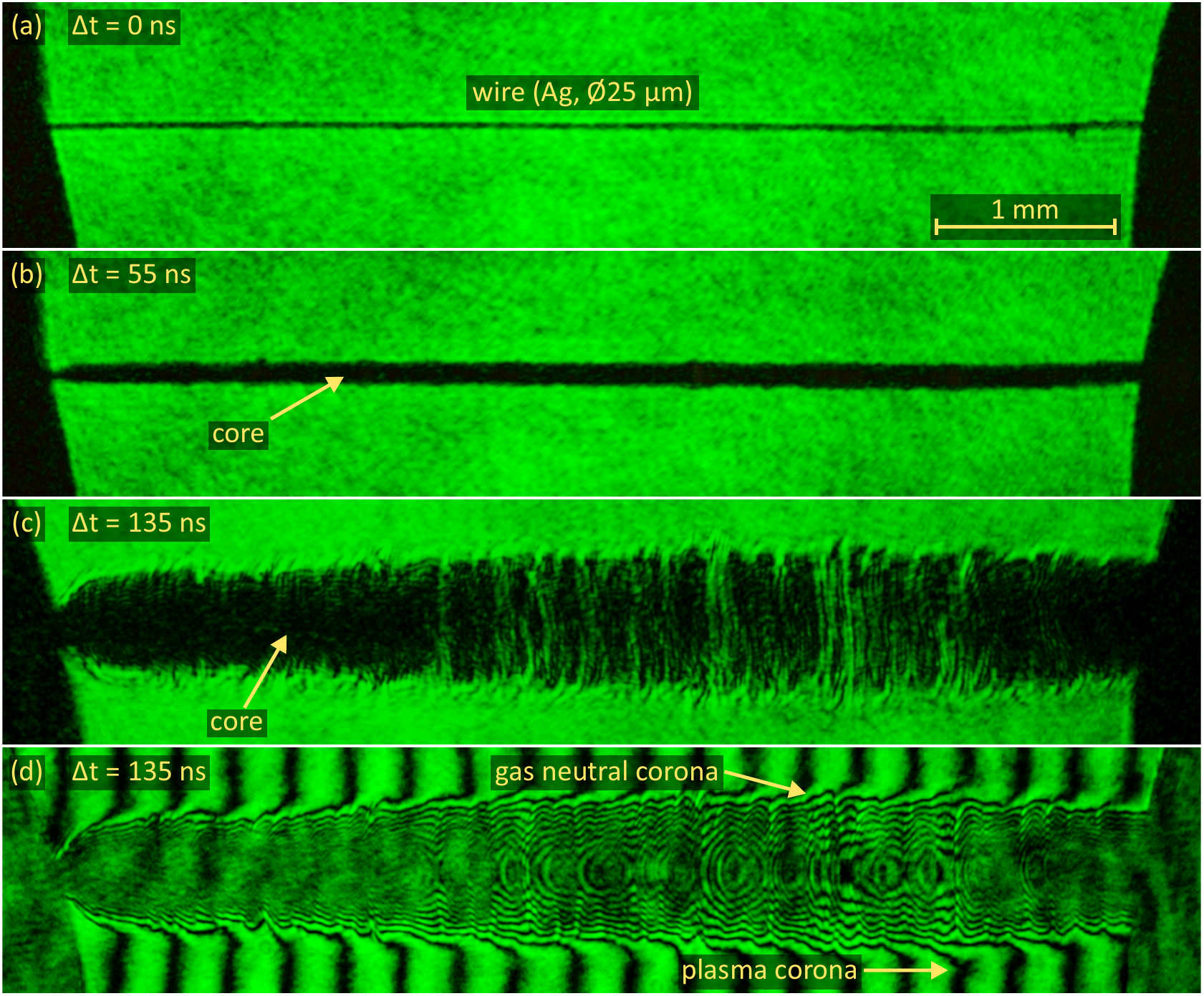}
\caption{\label{fig1} Single-shot shadowgrams (a,b,c) and interferogram (d) of the exploding silver wire registered with no applied current ($\Delta t = 0$~ns) and at different time points ($\Delta t = 55$~ns and 135~ns) of the explosion (shot no.~1). The cathode and anode are on the left and right sides of the images.}
\end{figure*}

Powerful tools for obtaining data on the explosion products generated during the EWE are laser probing techniques, in particular shadow imaging and interferometry \cite{guskov_2003, sarkisov_2004, shafer_2013, shi_2015, romanova_2018}. Basically, these diagnosing techniques rely on different methods of visualization of the changes in the intensity and phase of the laser radiation after its passage through the exploding wire. By analyzing the key mechanisms responsible for the changes in the radiation parameters, one can estimate the properties (phase state, density, matter composition, etc.) of the explosion products as well as reveal their structure. As an example, figure~\ref{fig1} shows typical shadowgrams and an interferogram of a silver wire (25~$\mu$m in diameter) exploded in vacuum ($\sim 10^{-5}$~Torr) by using a generator with charging voltage $U_0 \approx 20$~kV, maximum current $I_0 \approx 5$~kA, and current rise rate $dI/dt \approx 100$~A/ns \cite{romanova_2018}. The images were obtained via probing with a laser beam with a duration and wavelength of 70~ps and 0.532~$\mu$m at several time points ($\Delta t$) of the EWE, which are counted from the instant of the application of the current pulse to the load. Rapid resistive heating of the wire (Fig.~\ref{fig1}(a)) leads to its explosion and subsequent expansion of the explosion products. At the early stage, the exploded wire is visualized as an opaque cylindrical object with a sharp boundary, which is usually referred to as ``core'' (Fig.~\ref{fig1}(b)). During the expansion of the explosion products, the sharpness of the core boundary is vanished, and its surface is perturbed (Fig.~\ref{fig1}(c)). The core body becomes more transparent to the laser radiation and acquires a pronounced structure. Note that details of the explosion products visualized in shadowgrams greatly vary depending on the wire material and the initial parameters of the EWE \cite{romanova_2015}. An expanded core can appear to be homogeneous or have a more complicated structure with characteristic diffuse obscurations, jets, and gaps. The explosion products often turn out to be stratified in the transverse direction (this is also clearly seen in Fig.~\ref{fig1}(c)).

In some cases, visualization of the explosion products in shadowgrams is explained in terms of the light absorption in metal vapor (see e.g. in Ref.~\onlinecite{bilbao_2017}), even without taking into account the light refraction on gradients of the vapor density. In fact, the ``true'' absorption coefficient for vapors of most metals (constituting clusters of almost completely isolated atoms) at high pressure is very small and contains sharp maxima only within very narrow spectral bands (with the widths of few hundredths of an angstrom) \cite{Landsberg}. In this regard, the opacity of the core seen in Fig.~\ref{fig1}(c) to 0.532~$\mu$m laser radiation can hardly be accounted for by the absorption in an expanding cylindrical column of metal vapor far from resonance absorption lines. This is also the case for the light reflection (also away from absorption lines) from a uniform metal vapor column with a high density (not higher than $10^{22}$~cm$^{-3}$). Note that, in our case, core diameter $d$ in Fig.~\ref{fig1}(c) exceeds initial wire diameter $d_0 = 25$~$\mu$m by approximately thirty times, i.e. one can expect the average atomic concentration to be of $\sim 10^{20}$~cm$^{-3}$ at this moment (if the exploded wire is completely vaporized). Also, at the late stage of the EWE, a negligible contribution to the change in the intensity of the transmitted radiation is introduced by the light reflection and absorption by free electrons in plasma as well as by Rayleigh scattering by the equilibrium fluctuations of the vapor density. The corresponding analysis of the mentioned effects can be found in Ref.~\onlinecite{kolgatin_1989}.

It is more probable that the explosion products in shadowgrams are visualized due to multiple scattering of laser radiation by droplets with submicron sizes which are mixed with metal vapor (referred to as ``droplets in vapor'' \cite{vorobev_1997, tkachenko_2009}). Without delving into the nature of the droplets (theoretically, droplets constitute dense condensed matter), scattering by a single spherical droplet with diameter $a \lesssim \lambda$ can be characterized by scattering cross-section $\sigma_0 \sim 4/3 (2\pi)^5 a^6 / \lambda^4$, see in Ref.~\onlinecite{hulst_1958}. In the limiting case (when the exploded wire is assumed to be completely fragmented into the droplets), number $\tilde n$ of droplets per unit volume can be estimated as $(4/3 \pi a^3)^{-1} (d_0 / d)^2$, where $d_0$ and $d$ are the wire diameters before the explosion and at the instant of probing. Let us introduce characteristic attenuation coefficient $\chi$ which relates intensity $I_0$ of the incident radiation to intensity $I$ of the transmitted radiation as $I = I_0 \exp(-\chi L)$. Parameter $L$ characterizes the ``ray path'' in the matter. The attenuation coefficient can be expressed as $\chi = \tilde n \sigma_0$. For the exploded wire in Fig.~\ref{fig1}(c), taking $L$ to be of $\sim d$ in the limiting case, the value of $\chi L$ can be estimated as $\sim A (a / \lambda)^3$, where $A \sim 5 \times 10^3$ is the dimensionless factor obtained with $\lambda = 0.532$~$\mu$m, $d_0 = 25$~$\mu$m, and $d = 30 d_0$. It is seen that, with $a / \lambda \sim 0.1 \textrm{--} 1$, the effect of the light scattering can lead to significant changes in the intensity of the radiation passing through the cluster of submicron droplets.

Additional information about the state and structure of the explosion products can be extracted from the interferogram in Fig.~\ref{fig1}(d) obtained simultaneously with the shadowgram. By analyzing the fringe deflection, the regions with explosion products having different refractive indexes can be qualitatively distinguished. In particular, there is a region corresponding to a ``gas neutral corona'' which constitutes a $\sim 50$~$\mu$m thick layer of the exploded wire material near the core surface and has refractive index $n > 1$. In turn, the neutral corona is surrounded by a current-carrying plasma shell (``plasma corona'') with a thickness of $\sim 20 \text{--} 30$~$\mu$m and $n < 1$. In this region, fringes in the interferogram are deflected in the direction opposite to that characterizing the fringe deflection in the region with the neutral corona. The matter of the neutral corona introduces a noticeable phase shift (almost two units of the fringe shift), which indicates the presence of the high-density matter. However, in contrast to the core, some zones of the neutral corona remain relatively transparent to the laser radiation and probably consist mainly of metal vapor. At the same time, the fringe contrast sharply drops when approaching the core boundary. It can be assumed that here the effect of the multiple scattering by the submicron droplets also dramatically increases but verification of this fact in the case of interferograms requires a more rigorous consideration. Yet, looking ahead, note that the results of soft x-ray radiography actually confirm the complicated composition of the mater formed from the exploded wire (see also in Refs.~\onlinecite{guskov_2003, pikuz_1999}) and correlate with the assumption about submicron droplets originated during the EWE.

\section{\label{sec3}Two-wavelength laser probing}

\begin{figure}
\includegraphics [width=8.4cm] {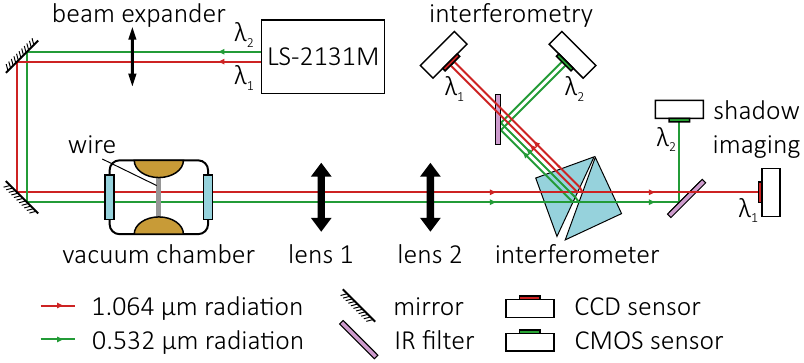}
\caption{\label{fig2} Schematic representation of the optical setup with simultaneous laser interferometry and shadow imaging at two wavelengths.}
\end{figure}

Remarkably, the scattering cross-section depends on the wavelength as $\sigma_0 \sim \lambda^{-4}$. If the multiple scattering by submicron droplets does occur in the experiments, then the observed pattern of the explosion products should be dramatically changed in the images when varying $\lambda$. For direct verification of this assumption, we carried out the experiment where the exploded wire was probed with the laser radiation at two wavelengths. To this end, we used the generator described in detail in Ref.~\onlinecite{ter-oganesyan_2005}. The generator has charging voltage $U_0$ and maximum current $I_0$ (supplied to the load) of 20~kV and 10~kA with the current rise rate being $\sim 50$~A/ns. To implement the laser probing techniques, we created an optical system (Fig.~\ref{fig2}) with its optical channels allowing simultaneous registration of shadowgrams and interferograms at two wavelengths (1.064~$\mu$m and 0.532~$\mu$m). Herein an Nd:~YAG laser (Lotis LS-2131M), which provides the beam with the energy and pulse duration of 120~mJ (for both harmonics) and 7--9~ns, was used. The achieved synchronization accuracy between the generator and the laser was better than $\sim 10$~ns. A photodetector (Thorlabs DET10A/M) with a rise time of 1~ns placed in front of a vacuum chamber was used to time the beam arrival. A standard Sigma 70--300~mm F4-5.6 DG Macro telescopic lens (lens~1 in Fig.~\ref{fig2}) having the resolution better than $\sim 100$~lines/mm collected the probing beam. Lens~1 was combined with a high-quality short-focus ``Era-13'' lens (lens~2 in Fig.~\ref{fig2}; see the lens characteristics in Ref.~\onlinecite{Volosov_1978}) having the resolution of 350~lines/mm. The beam transmitted through the lenses was divided into beams (with both harmonics) used in optical channels for interferometry and shadow imaging. Glass IR filters were introduced into the channels for selecting the beams with different wavelengths by means of transmitting the 1.064~$\mu$m and reflecting (from the front side of the filter) 0.532~$\mu$m light. Interferograms were obtained by using the beams reflected from the front and back sides of an air wedge between two rectangular glass prisms of a shearing interferometer \cite{pikuz_2001, parkevich_OLE_2019}, whereas shadowgrams were registered by using the beam transmitted through the interferometer. The achieved resultant spatial resolution of the imaging system was $\sim 10$~$\mu$m for each optical channel. The images were recorded by digital cameras (at 0.532~$\mu$m, by Canon EOS 1100D with CMOS sensor and, at 1.064~$\mu$m, by Nikon D80 with CCD sensor and removed IR cut-off filters) coupled to green and neutral glass filters for attenuating the self radiation of the exploding wire.

\begin{figure*}[t!]
\includegraphics [width=15.6cm] {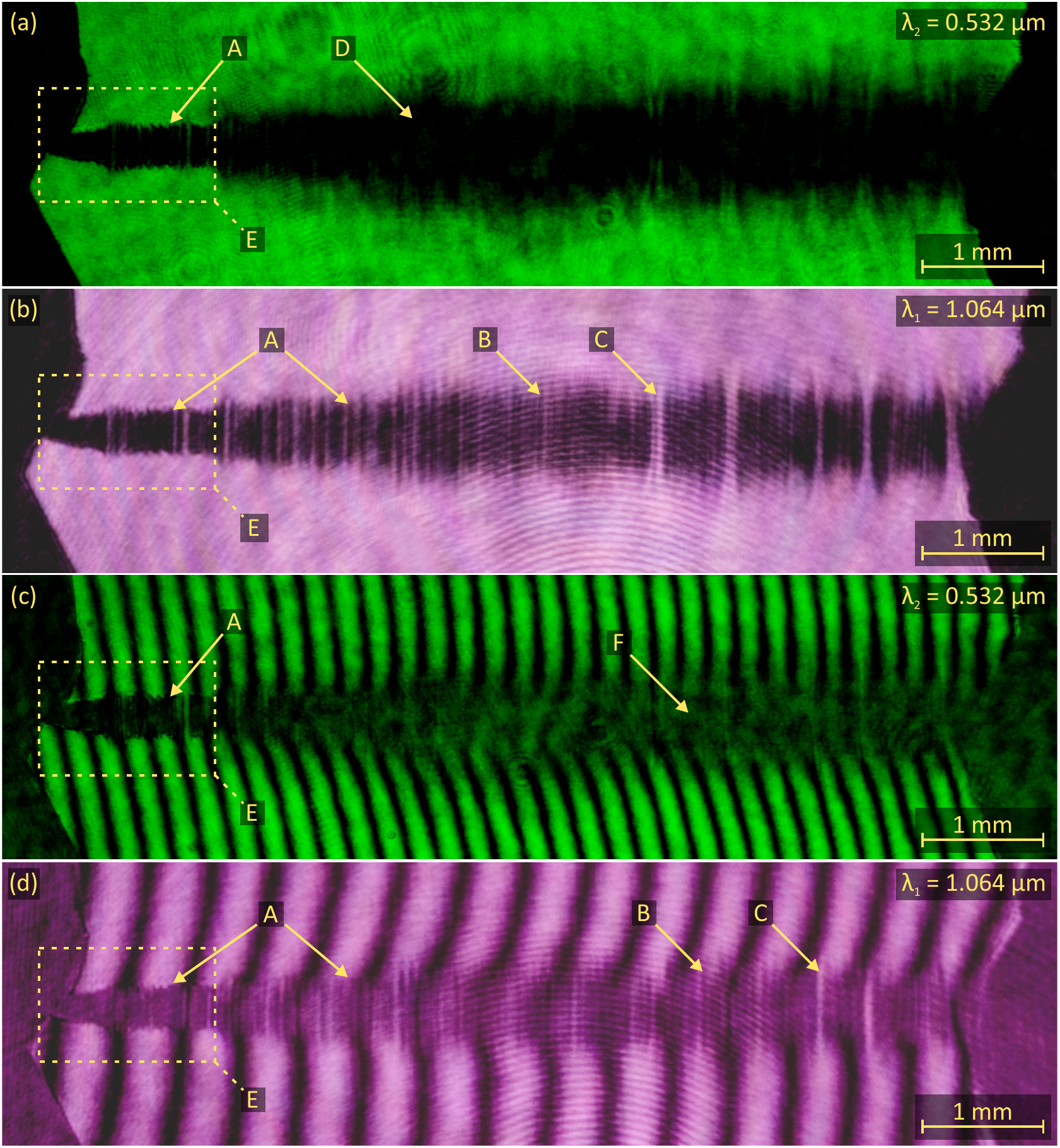}
\caption{\label{fig3} Single-shot shadowgrams (a,b) and interferograms (c,d) of the exploding gold wire (with 25~$\mu$m diameter) registered at two wavelengths (shot no.~2). The images are obtained 850~ns after ($\Delta t$) the application of the current pulse; A---core, B---striations, C---gaps, D---diffuse cloud, E---near-cathode region, F---regions with the sharp drop in the fringe contrast. The cathode and anode are on the left and right sides of the images.}
\end{figure*}

Figure~\ref{fig3} demonstrates the shadowgrams (registered at two wavelengths) of the gold wire with a diameter of 25~$\mu$m and a length of $\approx 8$~mm exploded in vacuum ($\sim 10^{-5}$~Torr). The differences between the shadowgrams are obvious. The stratified core (denoted as A), which is well resolved in Fig.~\ref{fig3}(b), is only partially visible in Fig.~\ref{fig3}(a). It is clearly seen in Fig.~\ref{fig3}(b) that the core consists of tightly packed striations and gaps (denoted as B and C) with characteristic longitudinal sizes of $\sim 40 \textrm{--} 90$~$\mu$m, whereas in Fig.~\ref{fig3}(a) the core is surrounded by a diffuse cloud (denoted as D) of ``some substance'' opaque to 0.532~$\mu$m radiation. This cloud also fills the gaps between the striations thereby masking the ``true core'' in the shadowgram. Additionally, the cloud is relatively inhomogeneous and has local regions with weak transmission at 0.532~$\mu$m. Conversely, the cloud is completely transparent at 1.064~$\mu$m. Proceeding from relation $\sigma_0 \sim \lambda^{-4}$, one can assume that the observed cloud constitutes a complicated mixture formed by submicron droplets and metal vapor, and the cloud is visualized due to the multiple scattering. Meanwhile, visualization of the stratified core in shadowgrams is practically independent of $\lambda$. The core is opaque to both harmonics (as well as to soft x-ray radiation \cite{guskov_2003, pikuz_1999}), which points to the high density of the core matter. Note that the cloud covers most of the core except for the near-cathode region (denoted as E); here the differences between the shadowgrams are minimal. This fact can be directly connected with the features of the energy input and transformation of the matter in the near-cathode region \cite{duselis_2004}.

Figures~\ref{fig3}(c),(d) show interferograms obtained simultaneously with shadowgrams. In both interferograms, the fringe deflection points to the prevalence of the neutral ($n > 1$) component of the explosion products. Deflection of fringes in the opposite direction is not observed even when probing at 1.064~$\mu$m (fringe shift is proportional to $\lambda$ for the electronic component of the refractive index \cite{Zaidel_1997}). Notably, here we deal with the late stage of the EWE when the plasma corona occupies such a large space volume (e.g. see the results of Ref.~\onlinecite{tkachenko_2009_PPR}) that the phase shift acquired by the laser radiation in this region is below the sensitivity limit of single-pass interferometry. The presence of the diffuse cloud surrounding the core leads to the sharp drop in the fringe contrast in the interference pattern in Fig.~\ref{fig3}(c) but, at the same time, does not affect the fringe quality (except for the zones containing the striations) in Fig.~\ref{fig3}(d) along the entire exploded wire. The drop in the fringe contrast in Fig.~\ref{fig3}(c) can be ascribed to the strong reduction in the spatial coherence of the transmitted radiation due to the multiple scattering by the submicron droplets randomly distributed in space. Additionally, we note that some parts of fringes in Figs.~\ref{fig3}(c),(d) are visualized in the region of the opaque core. This can be explained by the existence of ``transparency'' regions, which allow local transmission of the laser radiation while partially preserving its spatial coherence. This fact is probably the result of a combination of the following circumstances. First, the fringe step in Figs.~\ref{fig3}(c)(d) is much greater than the characteristic longitudinal sizes of the gaps and striations. Second, the cloud surrounding the core is itself inhomogeneous, i.e. it allows partial transmission of 0.532~$\mu$m radiation in local areas. Thus, when combined, the multiple small-scale zones of partial visualization of a single fringe in the core region provide visualization of the entire fringe (see Figs.~\ref{fig3}(c),(d)).

\section{\label{sec4}Problem of droplet origination and their properties}

\begin{figure}
\includegraphics{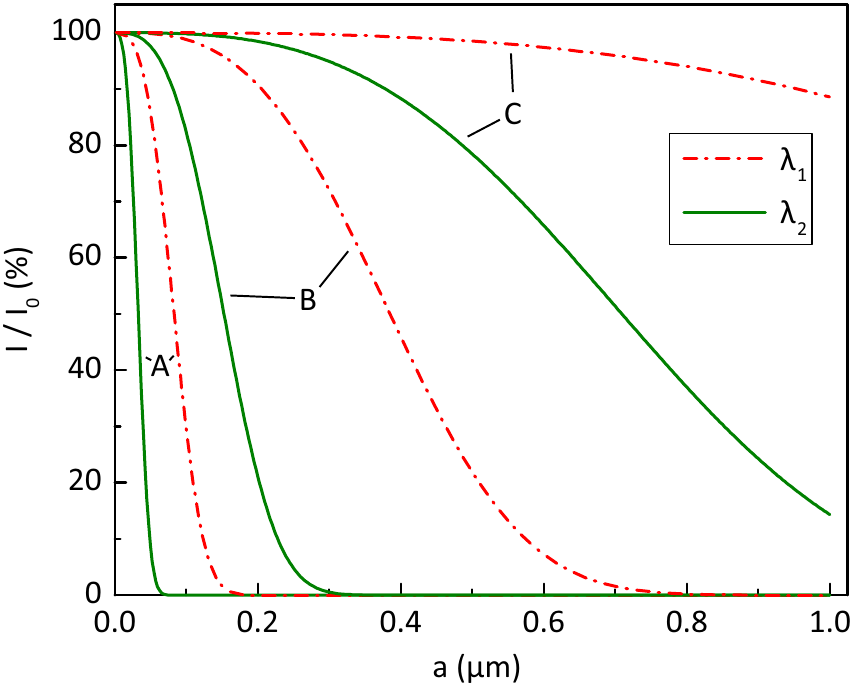}
\caption{\label{fig4} Intensity attenuation curves $I / I_0 \approx \exp(-\chi d)$ computed for $\lambda_1 = 1.064$~$\mu$m and $\lambda_2 = 0.532$~$\mu$m as functions of droplet size $a$ and density $\tilde n$: A---10$^{11}$~cm$^{-3}$, B---10$^9$~cm$^{-3}$, C---10$^7$~cm$^{-3}$. Parameter $d$ is taken as $\approx 50 d_0$.}
\end{figure}

It should be noted that to better illustrate the complicated composition of the explosion products generated during the EWE, we intentionally chose an example of the wire explosion (Fig.~\ref{fig3}) with a large energy deposition, as high as 4.4~eV/atom. This value exceeds the atomization energy of gold, so the degree of conversion of the wire material into metal vapor, in this case, is quite high. In spite of this fact, our findings indicate the presence of a significant fraction of the initial wire material, which is probably a complex mixture of metal vapor and submicron droplets. It can be assumed that the droplets are formed at the instant of phase transitions in the wire material occurring at the early stage of the explosion. The metastable superheated liquid, which appears at the beginning of the phase transitions in the outer part (where the magnetic field is high) of the expanding core, turns into supersaturated vapor (``droplets in vapor'' state). Later, when an unloading wave approaches the core axis and is reflected from it, a stretched liquid is formed which crushes itself during rapid expansion. In both processes associated with the so-called ``phase- and cavitation-type explosions'' \cite{romanova_2018} that occur in the outer and inner parts of the exploding wire, the generated explosion products contain condensed phase fragments (metallic or dielectric liquid which subsequently transforms into the solid phase). These fragments (in metallic or dielectric states) are specific for each part (outer or inner) of the exploding wire. As the energy input into the core by the transmitted current is increased, droplets are randomly distributed over the entire volume of the exploding wire. The droplet sizes can range from few nanometers to up to $\sim 1$~$\mu$m (characteristic scale of the metal crystallites). 
However, knowing the degree of the cloud opacity (much less than $\sim 10\%$ in Fig.~\ref{fig3}(a)) and the cloud transparency (more than $\sim 90 \%$ in Fig.~\ref{fig3}(b)), one can expect that the key role in the cloud visualization belongs to the droplets with the sizes smaller than 1~$\mu$m but, certainly, greater than tens of nanometers, with the characteristic density being $\tilde n \sim 10^8 \textrm{--} 10^{10}$~cm$^{-3}$. This is indicated by the intensity attenuation curves computed for both harmonics, see Fig.~\ref{fig4}. The curves are obtained for several cases of the droplet density, starting from the maximum possible ($\tilde n \sim 10^{11}$~cm$^{-3}$ for $d \approx 50 d_0$, $d$ is the cloud diameter in Fig.~\ref{fig3}(a)). Actually, droplets originated during the EWE have different sizes but the largest contribution to the intensity attenuation at 0.532~$\mu$m is made by those having the sizes of about several hundreds of nanometers. Unfortunately, the modern EWE physics gives little information about the discussed problem as well as about the phase state of the droplets at the instant of their observation; namely, it is unclear whether they are in the dielectric or metal state. So, we believe that only future studies of this issue may reveal the key properties of the droplets originated at one or another stage of the EWE.

\section{\label{sec5}Remark on dense core}

\begin{figure*}
\includegraphics {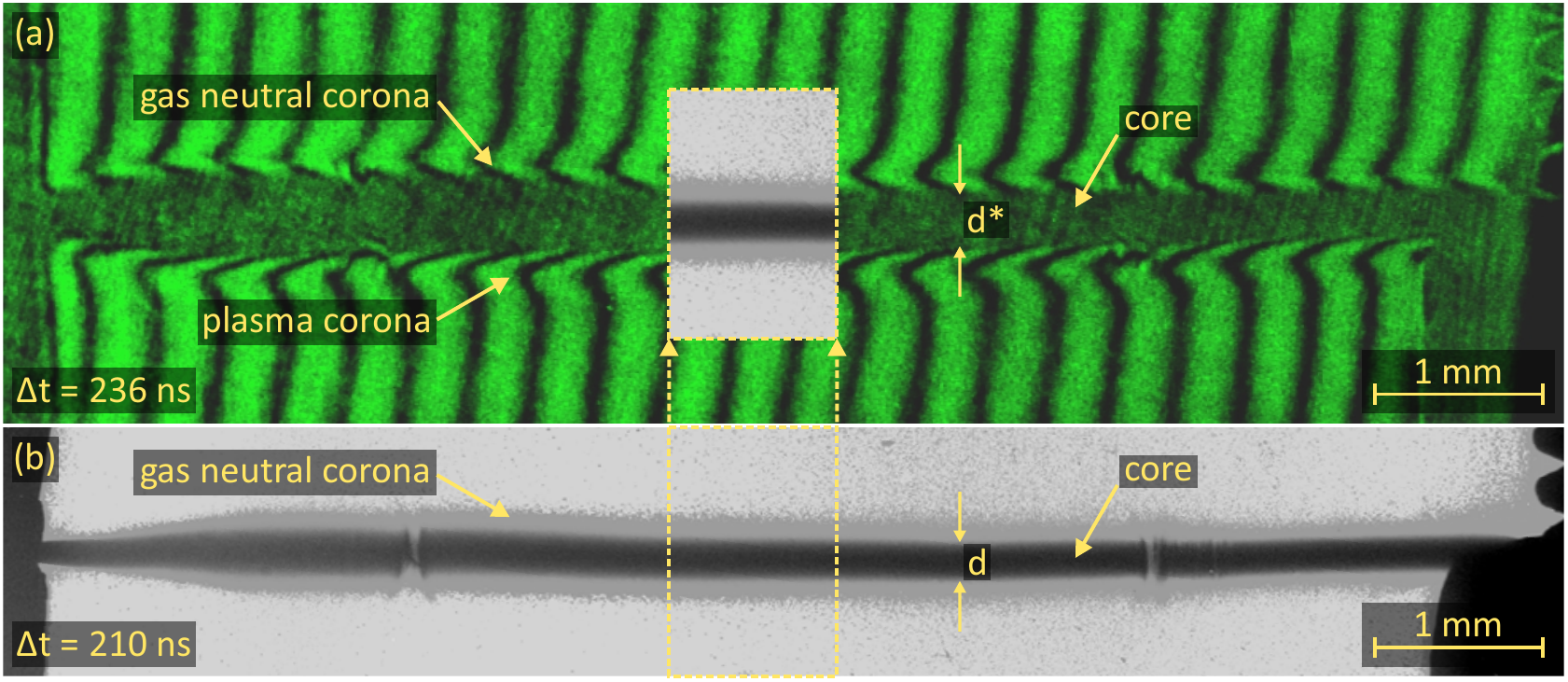}
\caption{\label{fig5} Single-shot interferogram (a) and the x-ray image (b) of the exploding gold wire (with a diameter of 20~$\mu$m) registered 236~ns and 210~ns after ($\Delta t$) the application of the current pulse (shot no.~3). The cathode and anode are on the left and right sides of the images.}
\end{figure*}

When studying the expansion of the explosion products during the EWE, it is necessary to determine the characteristic boundaries of the exploding wire. Typically, as widely done by numerous research groups, the boundaries of the exploding wire are attributed to the opacity zone in shadowgrams or interferograms. Notably, this zone is also assumed to be associated with the dense core. This would be a consistent assumption provided that the cloud of the submicron droplets hiding the ``true core'' is disregarded. The problem of the dense core is that this part of the exploding wire has a complicated structure visualization of which is of an ambiguous character. The case in point are the shadowgrams in Figs.~\ref{fig3}(a),(b) as well as the results of soft x-ray radiography combined with laser interferometry shown in Fig.~\ref{fig5}, which presents the exploded gold wire with a diameter of 20~$\mu$m. The shot was made by using a generator with $U_0 \approx 15$~kV, $I_0 \approx 4$~kA, and $dI / dt \approx 12$~kA/ns \cite{shelkovenko_2007}. It is seen that diameter $d$ of the true dense core in the x-ray image is several times smaller than diameter $d^*$ of the core observed in the interferogram. By matching the regions with different matter compositions, one can conclude that a significant part of the opaque core in Fig.~\ref{fig5}(a) is in fact filled with low-density matter of the gas neutral corona surrounding the dense core in Fig.~\ref{fig5}(b). The neutral corona coincides with the region of the sharp drop in the fringe contrast in Fig.~\ref{fig5}(a). By analogy with the shots considered above, it can be assumed that some zones of the neutral corona appear as a complicated mixture of metal vapor and submicron droplets, and their opacity is driven by multiple scattering. So, the submicron droplets significantly affect the visualization of the explosion products by laser probing. Hence, these droplets should be taken into account for the analysis of the corresponding images.

\section{\label{sec6}Outlook}

Multiple scattering turns out to be important when measuring the dynamic dipole polarizability, especially in the case of atoms of those metals that are rather difficult to obtain in the completely gaseous state. There are a number of studies (see e.g. Ref.~\onlinecite{sarkisov_2019} and references therein) where the integrated-phase technique was employed to obtain the polarizabilities of some metals.For this technique to be valid, the exploded wire must be completely vaporized at least in some region of the exploded wire. The existence of the submicron droplets mixed with metal vapor can challenge the obtained data on the polarizabilities, and, hence, this problem has to be rigorously analyzed in the future.

Note that reliable data on the explosion products are highly important for the development of a consistent model of the EWE. Alongside with natural fundamental aspects, such model is in demand for finding new methods for controlling the manufacturing of ultrafine powders \cite{kotov_2003}. Now, this process is largely empirical and based on the simple concept of conversion of the wire material into metal vapor with its further condensation into nanoparticle clusters.

Finally, apart from the EWE, the problem of the droplet origination is also related with the origination of plasma spots at electrodes during gas and vacuum discharges. For instance, when probing the near-electrode plasma formed after the breakdown in small air gaps, some plasma areas turn out to be opaque to 0.532~$\mu$m radiation \cite{parkevich_PSST_2018, parkevich_LR_2019, parkevich_2019_S}. Combined with the analytical considerations, the experimental facts point to the existence of dense near-electrode plasma jets generated in electrode local regions due to multiple explosions. The latter are caused by the high-power energy input into the electrode surface. The appearing droplets can significantly enhance the ionization of the surrounding air and, hence, promote the discharge development as well as accelerate its transition into the high-current regime.

Thus, the comprehensive investigation of the droplet origination and their properties can significantly advance the fundamental physics of condensed matter formed in different complex systems in extreme states.

\begin{acknowledgments}

This work was supported by the Russian Science Foundation under Grant 19-79-30086. Laser probing system development was supported in part by the National Nuclear Security Administration Stewardship Sciences Academic Programs through the Department of Energy under Grant No. DE-NA0003764. The image analysis was founded by the grant of the President of the Russian Federation (no.~MK-703.2020.2).

\end{acknowledgments}

\appendix

%\nocite{*}

\bibliography{LSSD}% Produces the bibliography via BibTeX.

\end{document}